\documentclass[aps, prl, reprint, superscriptaddress, floatfix, longbibliography, nobibnotes]{revtex4-2}

\usepackage[utf8]{inputenc}
\usepackage{amsmath}
\usepackage{amsfonts}
\usepackage{amssymb}
\usepackage{siunitx}
\usepackage{physics}
\usepackage{stmaryrd}
\usepackage[colorlinks=true, citecolor=blue, urlcolor=red]{hyperref}
\usepackage{orcidlink}

\newcommand{\mean}[1]{\langle {#1} \rangle}
\newcommand{\PI}{P_\mathrm{I}}
\newcommand{\PII}{P_\mathrm{II}}
\newcommand{\PIII}{P_\mathrm{III}}

\DeclareMathOperator{\erfc}{erfc}

\newcommand{\tfp}{t_\mathrm{fp}}
\newcommand{\e}{\mathrm{e}}
\newcommand{\trelax}{\tau_r}

\begin{document}
\title{First passage time for an underdamped harmonic oscillator and application to the power of an information engine}

\author{Aubin Archambault\,\orcidlink{0009-0002-3373-357X}}
\affiliation{\href{https://ror.org/02feahw73}{CNRS}, \href{https://ror.org/04zmssz18}{ENS de Lyon}, \href{https://ror.org/00w5ay796}{Laboratoire de Physique}, F-69342 Lyon, France}
\author{Caroline Crauste-Thibierge\,\orcidlink{0000-0001-5502-0445}}
\affiliation{\href{https://ror.org/02feahw73}{CNRS}, \href{https://ror.org/04zmssz18}{ENS de Lyon}, \href{https://ror.org/00w5ay796}{Laboratoire de Physique}, F-69342 Lyon, France}
\author{Alberto Imparato\,\orcidlink{0000-0002-7053-4732}}
\affiliation{Department of Physics, University of Trieste, Strada Costiera 11, 34151 Trieste, Italy}
\affiliation{ Istituto Nazionale di Fisica Nucleare, Trieste Section, Via Valerio 2, 34127 Trieste, Italy}
\author{Sergio Ciliberto\,\orcidlink{0000-0002-4366-6094}}
\affiliation{\href{https://ror.org/02feahw73}{CNRS}, \href{https://ror.org/04zmssz18}{ENS de Lyon}, \href{https://ror.org/00w5ay796}{Laboratoire de Physique}, F-69342 Lyon, France}
\author{Ludovic Bellon\,\orcidlink{0000-0002-2499-8106}}
\email{ludovic.bellon@ens-lyon.fr}
\affiliation{\href{https://ror.org/02feahw73}{CNRS}, \href{https://ror.org/04zmssz18}{ENS de Lyon}, \href{https://ror.org/00w5ay796}{Laboratoire de Physique}, F-69342 Lyon, France}

\date{\today}

\begin{abstract}
The distribution of the first passage time $\tfp$ for the position $x$ to overcome a threshold $x_B$ is calculated in an underdamped harmonic oscillator. The proof combines several approaches based on the determination of the eigenvalues of the Kramers differential operator for the intermediate and long time regimes and on a Hamiltonian approximation for the short times. The theoretical predictions are in excellent agreement with the results of an experiment on an underdamped micro-cantilever. The knowledge of the $\tfp$ distribution opens the way to several applications, among them the precise estimation of the power of information engines, which we have also experimentally checked. 
\end{abstract}

\maketitle
\section{Introduction}
The estimation of the first passage time (FPT) distribution is a widely studied problem because of the large number of applications in fundamental and applied research~\cite{Redner_2001, Redner_2013, Sekimoto_2010, Bray_2013}. Typical examples are the estimation of reaction rates in chemical processes~\cite{Hanggi_1990,Reuveni_2014}, of persistence properties in non--equilibrium systems~\cite{Bray_2013}, of the efficiency of search algorithms~\cite{Benichou_2011, Evans2021}, of the extreme records statistics in a time series~\cite{Majumdar_2008,Godreche2017}, of the power extraction in information engines~\cite{Archambault-2025}. FPT formalism also finds numerous application in other fields of research such as biology~\cite{Godec2017,Shin2019}, astrophysics~\cite{Chandrasek} and computer science~\cite{Majumdar2005}. 

Many theoretical~\cite{Redner_2001, Redner_2013, Sekimoto_2010, Bray_2013, Evans2011} and experimental studies~\cite{Besga_2020, Besga_2021, Faisant_2021, Roichman_2020, Roichman_2025} have been performed on systems described by overdamped Langevin equations and the FPT distributions have been computed and measured in these cases. For the underdamped systems where inertia plays a role the FPT probability has been studied, but due to the second derivative of the displacement with respect to time in the stochastic equations of motion, the process becomes non-Markovian, and computing the FPT has proved to be a daunting task. For example, for the simplest case of a free Brownian particle, also termed the random acceleration process, described by $\ddot x=\eta(t)$ (with $\eta(t)$ a white noise), the solution of the FPT probability remained elusive for a considerable span of time, and is now known in closed form only in the long time limit (see Ref.~\onlinecite{Bray_2013} for a review).

In this Letter we give new insight into this problem and present a solution of the FPT distribution for a Brownian harmonic oscillator, with complementary approaches to cover the full range of quality factor $Q$. The approach relies upon a Hamiltonian approximation for the short time regime whereas for long $t$ we reach Kramers' escape problem and use the energy diffusion formalism. We show that the predictions are in good agreement with the results of an experiment performed on an underdamped micro-cantilever at $Q=7$. The method is applied to the estimation of the power of an information engine, with again a good agreement between theory an experiment. 

In an accompanying paper~\cite{FPTuSHO_Article}, we give full details of the calculations and illustrate with numerical simulations the effect of the quality factor on the results, with a focus on the long time behavior.

\section{Trigger time distribution}

We consider a harmonic oscillator characterized by its stiffness $k$, angular resonance frequency $\omega_0$, viscous damping $\gamma$ and quality factor $Q=m\omega_0/\gamma$, in equilibrium with a thermostat at temperature $T$. We define the unit length as $\sigma=\sqrt{k_BT/k}$ (with $k_B$ the Boltzmann constant), the unit time as $\omega_0^{-1}$, and the unit of energy as $k_BT$. The harmonic potential is then $U(x)=\frac{1}{2}x^2$, and the Langevin equation describing the dynamics writes:
\begin{equation}\label{Lan:eq}
 \ddot x +\frac{1}{Q} \dot x + U'(x) = \sqrt{\frac{2}{Q}} \eta
\end{equation}
with $\eta$ a delta correlated noise of unit variance.

Starting from the equilibrium distribution of position, we are interested in computing the probability distribution $P(\tfp)$ of the first passage time $\tfp$ for $x$ to overcome a threshold $x_B$. This problem is useful to take into account inertia in the above mentioned FPT applications which have been studied only in overdamped systems~\cite{Hanggi_1990, Reuveni_2014, Bray_2013, Benichou_2011, Evans2021, Majumdar_2008, Godreche2017}, and will allow us to compute the power of an information engine with inertia~\cite{Archambault-2025} as an application example.
 
\begin{figure}[t]
	\centering
	\includegraphics{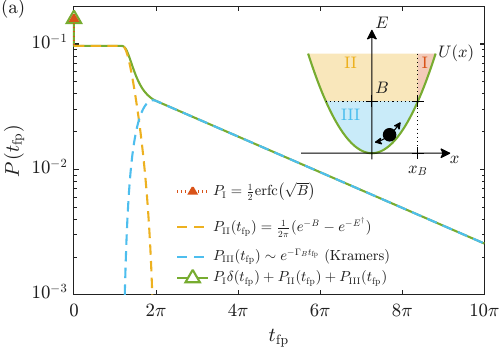}
 \includegraphics{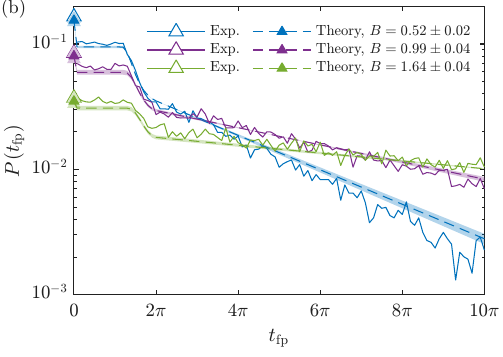}
	\caption{\label{Fig:pdftfp} (a) Theoretical probability distribution function (pdf) of the first passage time $\tfp$ for a harmonic oscillator of quality factor $Q=7$ to cross the threshold $x_B=1$. The pdf $P(\tfp)$ (green line) is the sum of three contributions (Eq. \ref{Eq:total_pdf}): $\PI$ (red vertical line ended by a triangle), $\PII$ (orange dashed line) and $\PIII$ (light blue dashed line). The inset pictures an underdamped particle moving in the harmonic potential $U(x)$ and the areas of initial values of position $x$ and total energy $E$ corresponding to the three contributions to $P(\tfp)$, with corresponding colors. Specifically: $\PI$ is a Dirac delta at $\tfp=0$ corresponding to instantaneous trigger ($x(t=0)\ge x_B$, area I in the inset); $\PII$ has a plateau followed by a quick decay corresponding to the remaining initial energies $E\ge B =\frac{1}{2}x_B^2$ (area II in the inset); $\PIII$ has an exponential tail corresponding to the initial energies $E<B$ (area III in the inset), escaping at rate $\Gamma_B$. (b) Distribution of $\tfp$ for three different values of the energy threshold $B$, demonstrating the very good agreement between experiment (plain line) and model (dashed line with the shaded area corresponding to the experimental uncertainty on $B$). }
\end{figure}

We decompose $P(\tfp)$ as the sum of three probability density functions (pdf):
	\begin{equation}
		P(\tfp)=\PI\delta(\tfp)+\PII(\tfp)+\PIII(\tfp), \label{Eq:total_pdf}
	\end{equation} 
where $\PI\delta(\tfp)$ corresponds to instantaneous trigger at $\tfp=0$, $\PII$ describes the first oscillation period $0< \tfp \le 2\pi$ and $\PIII(\tfp)$ captures the long time behavior. The contribution of the three pdfs is clearly seen in Fig.~\ref{Fig:pdftfp} where we plot $P(\tfp)$ at specific values of the energy barrier $B=\frac{1}{2}x_B^2$ and at $Q=7$. This functional form of $P(\tfp)$ has been tested by performing an experiment on an underdamped micro-lever and the experimental results, plotted in Fig.~\ref{Fig:pdftfp}(b), are in good agreement with the theoretical predictions.

\begin{figure}[t]
\centering
\includegraphics{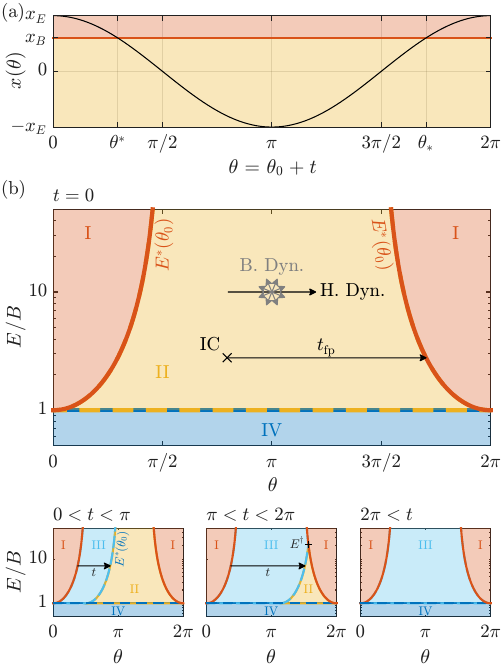}
\caption{\label{Fig:PhaseSpaceEt} (a) In the Hamiltonian dynamics approximation, the position is sinusoidal with an amplitude $x_E=\sqrt{2E}$, and can cross the threshold $x_B=\sqrt{2B}$ if $x_E\ge x_B$. It can happen instantaneously if the initial phase $\theta_0$ is smaller than $\theta^*(E)$ or larger than $\theta_*(E)$, and requires a time $\tfp=\theta_*(E)-\theta_0$ otherwise. (b) Phase space $(\theta,E)$ of the harmonic oscillator, at four different times (top left corner label). The red curves $E^*(\theta_0)$ delimit area I (red shaded), where $\tfp=0$, from area II (yellow shaded), where the Hamiltonian dynamics drives all initial conditions (IC) from left to right, and quickly cross the boundary of area I after a time $\tfp$ when $E>B$. Area II is upper bounded by a maximum energy $E^\dagger(t)$ when $\pi<t<2\pi$. Area IV (blue shaded) can only cross the barrier after diffusing due to Brownian noise into area III (light blue shaded), which replaces II during the first oscillation period $2\pi$. Movies of the phase space dynamics from direct numerical simulations of the Langevin equation \eqref{Lan:eq} are provided as ancillary files~\cite{SuppMatMovies} to support this picture.}
\end{figure}

Let us analyze the three contributions to $P(\tfp)$, starting with the short time behavior: $0 \le \tfp \le 2\pi$. At low damping ($Q\gg 1$), the total resonator energy $E=\frac{1}{2}\dot x^2+U(x)$ remains almost constant within a period. We can therefore use a Hamiltonian approximation in which the oscillator dynamics is well described by $x(t)= x_E \cos\theta$, where $x_E=\sqrt{2E}$, $\theta=\theta_0+t$ and $\theta_0$ sets the initial phase. In Fig.~\ref{Fig:PhaseSpaceEt}(a) $x$ is plotted as a function of $\theta$ and we clearly see that $x$ can overcome the threshold only if $x_E\ge x_B=\sqrt{2B}$, which implies that the initial energy satisfies $E\ge B$.

Instantaneous barrier crossing occurs if the initial position satisfies the condition $x(0)\ge x_B$. As the system is initially in equilibrium, using the Boltzmann distribution for the position we obtain:
\begin{equation} \label{Eq:Ptfpeq0}
	\PI=\int_{x_B}^\infty \frac{1}{\sqrt{2\pi}} \e^{-U(x)} dx = \frac{1}{2}\erfc\left(\sqrt{B}\right).
\end{equation}
In terms of initial phase $\theta_0$, this instantaneous trigger corresponds to the red area in Fig.~\ref{Fig:PhaseSpaceEt}:
\begin{equation} \label{Eq:theta*}
\text{or }\begin{cases}
 	\theta_0 \le \theta^*(E)= \arccos\left(\sqrt{\frac{B}{E}}\right)\\
 \theta_0\ge \theta_*(E)=2\pi-\theta^*(E).
\end{cases}
\end{equation}
In terms of initial energy, it translates into the condition 
\begin{equation}\label{Eq:ineq_theta}
 	E\ge E^*(\theta_0) = \frac{B}{\cos^2\theta_0} \text { and } \cos \theta_0>0.
\end{equation}
The set of initial conditions satisfying Eq.~\ref{Eq:ineq_theta} lie inside the red shaded areas I in the $(E,\theta)$ phase space depicted in Fig.~\ref{Fig:PhaseSpaceEt}(b), they are lower bounded by the red curve $E^*(\theta)$.

For all initial conditions outside the range of Eq.~\ref{Eq:ineq_theta} (yellow shaded area II in Fig.~\ref{Fig:PhaseSpaceEt}) a time
\begin{equation} \label{Eq:tfp}
\tfp=\theta_*(E)-\theta_0
\end{equation} 
is required to reach the threshold. Thus the dynamics within the first period is fully determined by $\theta_0$ and $E$, which are random variables. In the $(\theta,E)$ phase space, the initial probability distribution function is $P(\theta_0,E)=\frac{1}{2\pi}\e^{-E}$. Therefore, when $E\ge B$, the pdf of $\tfp$ in the range $[0,2\pi]$ is:
\begin{equation}
\PII(\tfp)= \int_{B}^\infty dE \ \psi(\tfp,E), \label{Eq:PtfpforEgtB}
\end{equation} 
with
\begin{equation}\label{Eq:PsiE}
\psi(E,\tfp)= \int_{\theta^*(E)}^{\theta_*(E)} d\theta_0 P(\theta_0,E)\delta(\theta_*(E)-\theta_0-\tfp).
\end{equation}
From Fig.~\ref{Fig:PhaseSpaceEt}, we notice that $\theta_0$ lies inside the interval $[\theta^*(E),\theta_*(E)]$. This implies a joint constraint on $\tfp$ and $E$ for the $\delta$-function in the integral of Eq.~\ref{Eq:PsiE} not to be $0$:
\begin{align}
	0 \le \tfp \le \theta_*(E)-\theta^*(E) &= 2\pi - 2\arccos\left(\sqrt{\frac{B}{E}}\right),\\
 \text{i.e. }-\cos(\tfp/2) & \le \sqrt{\frac{B}{E}}. \label{Eq:constrains}
\end{align}
In the integral of Eq.~\ref{Eq:PtfpforEgtB}, we can therefore replace the upper bound by $E^\dagger$, the maximum value for which the inequality \eqref{Eq:constrains} is satisfied: 
\begin{subequations} \label{Eq:Edagger}
	\begin{align} 
		E^\dagger(\tfp)&=\infty &&\text{if } 0<\tfp\le \pi, \label{Eq:Edaggera}\\
		E^\dagger(\tfp)&=\frac{B}{\cos^2(\tfp/2)} &&\text{if } \pi<\tfp\le 2\pi, \label{Eq:Edaggerb}\\
		E^\dagger(\tfp)&=B &&\text{if } 2\pi<\tfp. \label{Eq:Edaggerc}
	\end{align}
\end{subequations}
Eq.~\ref{Eq:Edaggerc} ensures that $\PII(\tfp)$ is zero for $\tfp>2\pi$ and will be useful to describe the case $E<B$ in the following. We now turn our attention back to Eqs.~\eqref{Eq:PtfpforEgtB}--\eqref{Eq:PsiE} and perform the integration over $\theta_0$ and $E$:
\begin{equation}\label{Eq:P_II}
	\PII(\tfp)= \frac{\e^{-B}-\e^{-E^\dagger(\tfp)}}{2\pi}.
\end{equation} 

When reaching $\tfp=2\pi$, area II has been emptied, leaving behind an empty space labeled III (light blue shaded area). All trajectory with initial conditions such that $E<B$ are trapped in area IV (dark blue shaded area labeled IV in Fig.~\ref{Fig:PhaseSpaceEt}), and would remain there forever for a purely Hamiltonian system. Actually, since the quality factor is finite, some diffusion occurs in $E$ and $\theta$ due to the Brownian forcing, and the frontiers between areas II, III and IV are porous. However, the density in the evolving phase space is stationary as long as the system is unaware of the presence of the threshold, and the contribution of areas I (Eq.~\ref{Eq:Ptfpeq0}) and II (Eq.~\ref{Eq:PtfpforEgtB}) to the pdf of $\tfp$ shouldn't depend strongly on $Q$. We first notice that initial microstates with $E<B$ occur with a probability
\begin{equation}
 P(E<B)=\int_0^B \int_0^{2\pi} P(\theta,E) d\theta dE=1-\e^{-B},
\end{equation}
and their initial energy is not high enough to reach the threshold $x_B$ (area IV). Thanks to the contact with a thermostat, the system can gain some energy, eventually explores area III by overcoming the barrier $B$ and finally crosses the threshold at $x_B$. This amounts to solving a Kramers escape problem, with the system that in principle can cross the energy barrier $B$ back and forth several times before reaching the target position $x_B$ with some non-negative velocity, so that the final energy at $x_B$ is $\frac{1}{2}(v^2+x_B^2)\ge B$.

In order to gain further insight into the first passage time distribution we now distinguish two cases: the limit of very large $Q$, and the finite $Q$ case.

When $Q\gg 1$, the first passage process across the position target $x_B$ amounts essentially to hit the energy $B$ for the first time, then to move on the circle with radius $B$ with an almost dissipationless dynamics toward the target at $x_B$. Thus for $Q\gg 1$ the first passage time through $x_B$ amounts to first passage time $\tau_Z(B)$ of an underdamped system across an energy barrier $B$, as discussed, e.g., by Zwanzig~\cite{Zwanzig-2001}. When $t<2\pi$, we note that energies accessible to the frontier between area III and I are lower bounded by $E^\dagger(t)$. The first passage time across this time dependent energy threshold is thus computed at all times using a barrier as $\tau_Z[E^\dagger(t)]$ (see Ref.~\onlinecite{FPTuSHO_Article} for details). As overcoming this threshold can occur at any phase during the oscillation, we then add $\pi$ (half a period) to the mean first passage time to account for the mean time to reach $x_B$ once the energy is large enough. The resulting escape rate is finally computed as
\begin{equation} \label{eq:GammaZwanzig}
 \Gamma(t)=\frac{1}{\pi+\tau_Z[E^\dagger(t)]}
\end{equation}
Once $\Gamma$ is known, it is straightforward to compute the pdf of first passage time corresponding to areas III-IV~\cite{FPTuSHO_Article}:
\begin{equation} \label{Eq:PIII}
 \PIII(\tfp)=(1-\e^{-B})\Gamma(\tfp)\exp\left(-\int_0^{\tfp} \Gamma(t)dt\right).
\end{equation}
Note that as soon as $t>2\pi$, $\Gamma=1/[\pi+\tau_Z(B)]\equiv\Gamma_B$ is constant and we simply expect an exponential tail for the distribution of $\tfp$.

For finite $Q$, reaching the energy $B$ is not sufficient to reach the target at $x_B$ within half a period, as the excitation is damped with a relaxation time $\trelax=2Q$, with the system returning in area IV several times before hitting $x_B$. In this case we can estimate the escape rate $\Gamma$ through direct evaluation of the slowest eigenvalue $-\lambda_1(x_B)$ of the differential Kramers operator associated with the stochastic equation \eqref{Lan:eq} with an absorbing condition in $x_B$. The contribution $\PIII(\tfp)$ to the first passage time pdf is then still evaluated with Eq.~\ref{Eq:PIII}, using $\Gamma(t)=\lambda_1[E^\dagger(t)]$. In this case also, $\Gamma(t)$ becomes time-independent for $t>2 \pi$ and the distribution displays an exponential tail of rate $\Gamma_B\equiv\lambda_1(x_B)$ (see ref.~\cite{FPTuSHO_Article} for the full proof).

By summing the contribution to the pdf of $\tfp$ corresponding to all regions of phase space (Eqs.~\ref{Eq:Ptfpeq0}, \ref{Eq:P_II} and \ref{Eq:PIII}), we have a full theoretical description of the distribution of the trigger time introduced by Eq.~\ref{Eq:total_pdf}. This pdf is parameterized by $B$ and $Q$ only. We present in Fig.~\ref{Fig:pdftfp}(a) the expected shape with the different contributions.

\section{Experiment}

In our experimental setup~\cite{Dago-2022-JStat,Dago-chapter,Archambault-2025}, the resonator is the fundamental vibration mode of micro-cantilever held at room temperature $T=\SI{300}{K}$. Its characteristics are: $\omega_0=2\pi\times \SI{1087}{Hz}$, $Q\simeq 7$, $k\simeq\SI{5e-3}{N/m}$, leading to $\sigma\simeq \SI{0.8}{nm}$. The deflection $x$ is measured with pm resolution with a quadrature phase differential interferometer~\cite{paolino_quadrature_2013} and proper thermal drift compensation. We set a threshold $x_B$, and starting from equilibrium, we record the position as a function of time at a sampling rate of $\SI{2}{MHz}$, far larger than the characteristic frequency response of the system. Once the threshold is reached (which can be instantaneous), we save the value of this first passage time $\tfp$ and perform an action (specifically extract work in Ref.~\onlinecite{Archambault-2025}). To restart from a fresh equilibrium state, we then wait a time $\tau=11\times2\pi\simeq 5\trelax$ before repeating the operation. For every choice of $x_B$, at least $\SI{200}{s}$ of data are acquired, corresponding to $\num{2e4}\,\tau$ and leading to a representative statistical sampling of the pdf of $\tfp$ (from $\sim\num{2e4}$ samples for $B=0$ to $\sim\num{3e3}$ samples for $B=4$).

\begin{figure}
\centering
\includegraphics{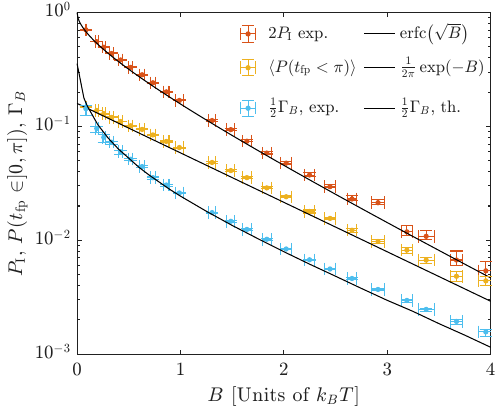}
\caption{\label{Fig:pdftfpcontributions} Main contributions to the pdf of $\tfp$ as a function of the energy threshold $B$: Probability $\PI$ of instantly overcoming the threshold $x_B$ (experimental data, red dots with error bar corresponding to one standard deviation of the statistical uncertainty, and model, black line); plateau of $P(\tfp)$ for $0<t<\pi$ (experimental data, yellow dots and model); Kramers' escape rate $\Gamma_B$ deduced from the exponential decay of $P(\tfp)$ for $t>2\pi$ (experimental data, blue dots, and model)~\cite{NoteNoCurveOverlap}. The theory-experiment agreement is very good.}
\end{figure}

In Fig.~\ref{Fig:pdftfp}(b), we plot a comparison of the theoretical model with the experimental data for a few values of $B$, showing the very good agreement between the two. To be more quantitative, we study in Fig.~\ref{Fig:pdftfpcontributions} the three contributions to the pdf: the probability of instant triggers ($\PI$), the value of the plateau of $\PII$ for $0<\tfp<\pi$, and finally the escape rate $\Gamma_B$ of the exponential tail ($\PIII\propto \e^{-\Gamma_B \tfp}$). The latter is simply evaluated from the slope of a linear fit of $\ln[P(\tfp)]$ for $\tfp>2\pi$. The model of each contribution accurately accounts for the experimental observations.

\section{Application: information engine power}
\begin{figure}[!ht]
\centering
\includegraphics{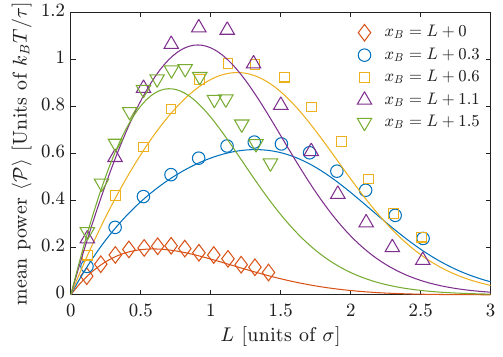}
\caption{\label{Fig:Power} Mean power of the information engine versus $L$ for various values of the threshold $x_B$. The theoretical expectation of Eq.~\ref{Eq:meanP} (plain line) is in reasonably good agreement with the experimental data (markers). For low values of $L$ and $x_B$, since the extracted work $\sim w_0$ is low, so does the power. In contrast, for large values of $L$ and $x_B$, $\mean{-w}$ is large, but the trigger rate drops as $\mean{\tfp}$ grows, resulting as well in a low engine power. The optimum lies in between for $x_B \sim 2L \sim 2$.}
\end{figure}

As an application of this first passage problem, we consider the information engine described in Ref.~\onlinecite{Archambault-2025}. This engine uses as a fuel the output of the continuous measurement that is started at equilibrium and performed till the oscillator position $x(t)$ reaches the target position $x_B$. At this instant $\tfp$, the center of the harmonic well is translated by a distance $2L$ from its initial position. The resulting work extracted during this operation is $w=2L(x(\tfp)-L)$. If $\tfp>0$, then the work is always $w_0=2L(x_B-L)$, but for instantaneous trigger ($\tfp=0$), it depends on the position $x(t=0)$ and is larger than $w_0$. We compute in Ref.~\onlinecite{Archambault-2025} that in average,
\begin{equation} \label{Eq:meanW}
\mean{w}=2L \left[x_B (1-P_I)+L\left(\frac{1}{\sqrt{2\pi}}\e^{-B}-1\right)\right].
\end{equation}
Now that we have the mean work, we need to compute the mean duration between two work extractions, which sums the mean first passage time $\mean{\tfp}$ and the dwell time $\tau=11\times2\pi\simeq 5\trelax$ we impose to start from a fresh equilibrium state before the next engine stroke. The mean power of the information engine is finally:
\begin{equation} \label{Eq:meanP}
 \mean{\mathcal{P}}=\frac{\mean{-w}}{\tau+\mean{\tfp}}.
\end{equation}
This expectation, plotted in Fig.~\ref{Fig:Power}, matches the experimental data. It also helps exploring the range of possible parameters for $x_B$ and $L$ to look for the maximum power. For our experimental choice of $\tau$, this optimum value is reached for $L=1.02$ and $x_B=2.03$. $L \sim x_B/2$ is expected since it maximizes $w_0$ for a given $x_B$, and $x_B\sim 2$ leads to $B\sim 2$, which is a compromise between \emph{large} and \emph{frequent} energy extraction.

\section{Conclusions}

To conclude, we have analytically computed the pdf of the first passage time for the position $x$ of an underdamped harmonic oscillator to reach a threshold $x_B$. The derivation is rather unusual and is based on a mechanistic approach from short to intermediate time, which can be applied because of the low dissipation (high $Q$). For long time we use the energy diffusion model to complete the full estimation of the pdf. 
Our result, which is given in closed form in eq.~\ref{Eq:total_pdf}, represents an advancement in the study of FPT statistics for underdamped systems, the only closed form solution for the pdf being the one for the acceleration process, i.e. for a free particle with $U(x)=0$, in the long time limit. 

We have tested the predicted functional form with that measured on the data of an experiment on an underdamped micro-cantilever, finding an excellent agreement between theory and experiments. We have also computed the power of an information engine fueled by the first passage time, finding again a good agreement with the experimental results.

Our theoretical result is very useful because it opens the way to the estimation of the first passage time in many applications in underdamped systems, in particular in the fields of stochastic and information thermodynamics, as discussed in this Letter. As a final note, this description of the full distribution of FPT is quite different from the Kramers rate theory, which is usually applied to diffusive crossing of large barriers. Indeed, in our distribution the exponential tail appears only for time much larger than one oscillation period. 

\acknowledgments
Data supporting this study will be available in an open public repository upon acceptance of the manuscript.\\
We thanks S. Majumdar and C. Jarzynski for enlightening discussions. This work has been partially funded by project ANR-22-CE42-0022. A.I. acknowledges support from the CNRS and the ENS de Lyon as an invited Researcher and invited Professor in the ENS de Lyon.
\bibliography{FPTuSHO}

\begin{thebibliography}{28}
\bibitem [{Not()}]{NoteNoCurveOverlap}%
  \BibitemOpen
  \href@noop {} {\bibinfo {title} {{Note that to avoid curve overlapping,
  $P_\mathrm{I}$ and $\Gamma_B$ have been offset by a factor $2$ and
  $\frac{1}{2}$ respectively, as noted in the legend.}}}\BibitemShut {Stop}%
\end{thebibliography}
\end{document}